\documentclass[useAMS,usenatbib]{mn2e}  
\usepackage{graphicx}
\usepackage{times}
\usepackage{amssymb} %%% for math symbols as $\lesssim$ and/or $\gtrsim$
\usepackage{txfonts}
\usepackage{appendix}
\usepackage{natbib}
\usepackage{longtable, portland}
\usepackage{supertabular}
\usepackage{rotating}
\usepackage{array}
\bibpunct{(}{)}{;}{a}{}{,}

\def\kms    {\ifmmode{{\rm ~km\,s}^{-1}}\else{~km\,s$^{-1}$}\fi}

\begin{document}

\title{The stellar mass--size relation for the most isolated galaxies in the local Universe.\footnotemark}

\author[Fern{\'a}ndez Lorenzo et al.]{M. Fern{\'a}ndez Lorenzo$^1$\thanks{E-mail:mirian@iaa.es}, J. Sulentic$^1$, L. Verdes--Montenegro$^1$, \and M. Argudo--Fern{\'a}ndez$^1$\\
$^1$Instituto de Astrof{\'i}sica de Andaluc{\'i}a, Granada,  IAA-CSIC Apdo. 3004, 18080 Granada, Spain \\
}

 \date{Received.....; accepted..... }

\maketitle

\label{firstpage}

\begin{abstract}

Disentangling processes governing the formation and evolution of galaxies is a fundamental challenge in extragalactic research. In this sense the current belief that galaxies grow by the action of minor mergers makes the study of the stellar mass--size relation in different environments an important tool for distinguishing effects of internal and external processes.
 
The aim of this work is to study the effects of environment on the growth in size of galaxies. As part of AMIGA project (Analysis of the Interstellar Medium of Isolated GAlaxies), we examine the stellar mass--size relation for a sample of the most isolated galaxies in the local Universe interpreted as stellar systems where evolution has been mainly governed by internal processes. Effects of environment on the stellar mass--size relation are evaluated by comparing our results with samples of less isolated early-- and late--type galaxies, as well as, for the first time, different spiral subtypes.

Stellar masses in our sample were derived by fitting the SED of each galaxy with {\tt kcorrect}. We used two different size estimators, the half--light radius obtained with {\tt SExtractor\rm} and the effective radius calculated by fitting a S\'ersic profile to the i--band image of each galaxy using {\tt GALFIT\rm}. We found good agreement between those size estimators when the S\'ersic index fell in the range 2.5$<$n$<$4.5 and 0.5$<$n$<$2.5 for (visually classified) early-- and late--type galaxies respectively.

We find no difference in the stellar mass--size relation for very isolated and less isolated early--type galaxies. We find that late--type isolated galaxies are $\sim$1.2 times larger than less isolated objects with similar mass. Isolated galaxies and comparison samples were divided into 6 morphological ranges (E/S0, Spirals, Sb, Sbc, Sc, and Scd--Sdm) and 5 stellar mass bins between $\log$(M$_{\ast}$)=[9,11.5]. In all cases the relation is better defined and has less scatter for the isolated galaxies. We find that as the morphological type becomes later the galaxy size (for a fixed stellar mass range) becomes larger. For the lowest stellar mass bins $\log$(M$_{\ast}$)=[9,10] we find good agreement between sizes of AMIGA and comparison spirals (both mostly composed of Scd--Sdm types). The isolated spiral galaxies in the high stellar mass bins $\log$(M$_{\ast}$)=[10,11] tend to be larger than less isolated galaxies. This difference in size is found for all spiral subtypes and becomes larger when we compare 
fully isolated galaxies with galaxies having 2 or more satellites (neighbors within 3 magnitudes of difference at a distance less than 250 kpc from the galaxy).

Our results suggest that massive spiral galaxies located in low density environments, both in terms of major companions and satellites, have larger sizes than samples of less isolated galaxies. Hence the environment has played a role in the growth in size of massive spiral galaxies.

\end{abstract}


\renewcommand{\thefootnote}{}\footnote{$^{\star}$Full table 1 is only available in electronic form at the CDS via anonymous ftp to cdsarc.u-strasbg.fr (130.79.128.5) or via http://cdsarc.u-strasbg.fr/viz-bin/qcat?J/MNRAS/vol/page, and at the AMIGA VO interface (http://amiga.iaa.es).}

\section{Introduction}

Several recent studies find evidence for growth in size of galaxies from redshift 2--3 to the present \citep{2005ApJ...626..680D, 2006ApJ...650...18T, 2007MNRAS.374..614L,2007MNRAS.382..109T,2008A&A...482...21C,2008ApJ...677L...5V}. Evidence is especially strong for the most massive early--type galaxies which have increased in size by a factor 4.3 since z$\sim$2.3. There is also evidence that late--type systems have increased their size 2.6 times over the same z interval \citep{2008ApJ...687L..61B}. Scenarios have been proposed to explain this size evolution including environmentally dependent and independent processes. The most likely and accepted mechanism involves growth from dry minor mergers \citep{2005ApJ...625...23B,2005AJ....130.2647V} belived to be more efficient for increasing the size than the stellar mass of galaxies. Consideration of this scenario and the fact that the number of compact objects is almost non--existent at z=0 \citep{2011A&A...526A..72F}, leads to the expectation of an increase 
with redshift in the number of galaxies with minor companions. However the fraction of galaxies with satellites appears to remain constant from z$\sim$2 to the present \citep{2012MNRAS.422.2187M}. Different mechanisms, depending on the galaxy mass, would be needed if secular processes are responsible for the growth in galaxy size. For the most massive galaxies, growth in size would be driven by quasar feedback which could remove huge amounts of cold gas from the central regions thereby inducing an expansion of the stellar distribution. On the other hand less massive galaxies would undergo adiabatic expansion as a consequence of mass loss driven by stellar winds and supernova explosions \citep{2008ApJ...689L.101F}. In both cases a significant evolution of the velocity dispersion (larger at high redshift) is predicted for both small and large galaxies. The evidence so far shows only mild evolution in velocity dispersion \citep{2009ApJ...696L..43C,2009ApJ...704L..34C,2011ApJ...738L..22M}.

If the observed differences are real and if they are due to environmental influences then comparisons at low redshift of the stellar mass--size relation for galaxies in different environments might shed some light on this interpretation. Previous environmental studies of the stellar mass--size relation were made by comparing field and cluster early--type galaxies. Most of them find no dependence of the stellar mass--size relation with environment at both local \citep{2010MNRAS.402..282M} and high redshift \citep{2010ApJ...709..512R,2010ApJ...721L..19V}. However, \citet{2008A&A...482...21C} found a different stellar mass--size relation for cluster and field galaxies at redshift $\sim$1. Cluster early--type galaxies seem to be located closer to the stellar mass--size relation at z=0 than those located in low density environments, which present smaller sizes at high--redshift. In addition \citet{2012MNRAS.419.3018C} found that early--type systems in higher density regions tend to be more extended than their 
counterparts in low--density enviroments. In the case of late--type galaxies \citet{2010MNRAS.402..282M} find evidence for environmental dependence on the stellar mass--size relation. However this evidence is marginal (2$\sigma$) and only for intermediate/low stellar masses. The most massive spirals follow the same stellar mass--size relation in field and cluster environments. They claim that there is a population of spirals containing extended disks only in the field.

Although minor mergers are the most popular explanation for the growth of galaxies in size, a dependence with environment is not clearly established, as we have seen before. Moreover, since the growth in size is stronger for massive early--type galaxies, the effect of environment should be larger for these objects. However, only spirals present a (weak) dependence with environment. Since field usually includes pairs and even groups of galaxies, interactions may have played an important role in the results found in previous studies. At this point a well defined environment can be crucial. 

While it is easy to recognize a rich cluster, definitions of low--density environments can be confusing. In recent years there has been an increased emphasis on identifying low density or isolated galaxy populations. One of the most useful samples remains the visually selected Catalog of Isolated Galaxies (CIG) compiled by \citet{1973AISAO...8....3K}, more recently vetted as the AMIGA sample \citep[Analysis of the interstellar Medium of Isolated GAlaxies,][and references therein]{2010ASPC..421....3S}. AMIGA galaxies show different physical properties than galaxies in denser environments (even field galaxies), including a lower infrared luminosity (L$_{FIR}$$<$ 10.5 L$\sun$) and colder dust temperature \citep{2007A&A...462..507L}, a low level of radio continuum emission dominated by mild disk star formation \citep{2008A&A...485..475L}, no radio active galactic nuclei (AGN) selected using the radio--far infrared correlation \citep[0$\%$;][]{2008A&A...486...73S} and a small number of optical AGN 
\citep[22$\%$;][]{2012A&A...545A..15S}, less molecular gas \citep{2011A&A...534A.102L}, or a smaller fraction of HI asymmetries \citep[$<$ 20$\%$,][]{2011A&A...532A.117E}. In addition, early--type galaxies in AMIGA are fainter than late types, and most AMIGA spirals host pseudo--bulges rather than classical bulges, as well as different level of optical asymmetry, clumpiness and concentration \citep{2008MNRAS.390..881D}. The comparison of colors between AMIGA and pairs of galaxies has shown a passive star formation and a gaussian distribution of colors for isolated galaxies, while the galaxies in pairs present bluer colors and higher color dispersions, which is indicative of a star formation enhanced by interactions \citep{2012A&A...540A..47F}. If environment is affecting the growth in size of galaxies, might isolated galaxies be smaller than other galaxies because they had undergone fewer minor mergers? \citet{2006A&A...456...91G} tried to answer this question by comparing the distribution in size between isolated and 
interacting galaxies. Their results suggest that isolated galaxies are smaller than objects in interaction, but a comparison of the size as function of the stellar mass was not made in that work, which can lead to a wrong conclusion if the two samples do not have the same stellar mass distribution. 

Here, we propose to analyze the stellar mass--size relation for the AMIGA sample of isolated galaxies. In Sect. 2, the sample selection and data analysis is described. The stellar mass--size relation is presented in Sect. 3, and the the discussion of the results is provided in Sect. 4. Finally, the conclusions are exposed in Sect. 5. Throughout this article, the concordance cosmology with ${\Omega}_{\rm \Lambda0}=0.7$, ${\Omega}_{\rm m0}=0.3$ and $\rm H_0=70 \rm \ km \rm \ s^{-1} \rm \ Mpc^{-1}$ is assumed for comparison with other studies about the stellar mass--size relation ($\rm H_0=75 \rm \ km \rm \ s^{-1} \rm \ Mpc^{-1}$ is assumed in the other papers of AMIGA). Unless otherwise specified, all magnitudes are given in the AB system.

\begin{table*}
\caption{Data for the AMIGA sample.}
\label{dat}
\begin{center}
\begin{tabular}{ccccccccccc}
\hline
\tiny
\\
\small
{\bf CIG} & {\bf Type} & {\bf Mag$_g$} & {\bf Mag$_r$} & {\bf Mag$_i$} & {\bf Mag$_{K_S}$} & {\bf $\log$ (M$_{\ast}$)} & {\bf R$_{50}$} & {\bf n} & {\bf R$_e$} & {\bf b/a} \\
{} & {\bf (RC3)} & {\bf (mag)} & {\bf (mag)} & {\bf (mag)} & {\bf (mag)} & {\bf (M$_{\odot}$)} & {\bf (kpc)} & {} & {\bf (kpc)} & {} \\
{\bf (1)}  & {\bf (2)}  &  {\bf (3)} & {\bf (4)} & {\bf (5)} & {\bf (6)} & {\bf (7)} & {\bf (8)} & {\bf (9)} & {\bf (10)} & {\bf (11)} \\
\hline
   2 &  6 & 14.83$\pm$0.04 & 14.37$\pm$0.03 & 14.11$\pm$0.02 & 14.10$\pm$0.12 & 10.18$\pm$0.05 &  4.72$\pm$0.14 &  1.09$\pm$0.01 &   4.64$\pm$0.02 &  0.59 \\
   4 &  3 & 12.22$\pm$0.01 & 11.47$\pm$0.01 & 11.04$\pm$0.01 & 10.24$\pm$0.01 & 10.63$\pm$0.01 &  4.03$\pm$0.03 &  0.67$\pm$0.00 &   3.35$\pm$0.00 &  0.26 \\
   5 &  0 & 15.24$\pm$0.04 & 14.53$\pm$0.03 & 14.16$\pm$0.03 & 13.66$\pm$0.04 & 10.45$\pm$0.03 &  2.95$\pm$0.10 &  1.14$\pm$0.01 &   2.66$\pm$0.01 &  0.38 \\
   6 &  7 & 14.28$\pm$0.03 & 13.89$\pm$0.02 & 13.70$\pm$0.02 & 13.64$\pm$0.10 &  9.88$\pm$0.05 &  2.97$\pm$0.08 &  1.72$\pm$0.01 &   2.99$\pm$0.02 &  0.43 \\
   7 &  4 & 14.75$\pm$0.03 & 14.12$\pm$0.02 & 13.79$\pm$0.02 & 13.29$\pm$0.07 & 11.00$\pm$0.03 &  7.28$\pm$0.16 &  1.92$\pm$0.01 &   7.88$\pm$0.05 &  0.70 \\
   9 &  5 & 14.84$\pm$0.03 & 14.36$\pm$0.03 & 14.07$\pm$0.02 & 13.61$\pm$0.07 & 10.43$\pm$0.04 &  5.17$\pm$0.19 &  1.00$\pm$0.01 &   4.23$\pm$0.02 &  0.28 \\
  12 &  3 & 15.17$\pm$0.04 & 14.61$\pm$0.03 & 14.34$\pm$0.03 & 13.64$\pm$0.09 &  9.95$\pm$0.04 &  2.47$\pm$0.11 &  1.23$\pm$0.01 &   2.15$\pm$0.01 &  0.31 \\
  13 & -5 & 13.55$\pm$0.03 & 12.82$\pm$0.02 & 12.42$\pm$0.01 & 12.04$\pm$0.04 & 10.78$\pm$0.02 &  2.24$\pm$0.03 &  5.24$\pm$0.01 &   3.21$\pm$0.01 &  0.76 \\
  14 & -3 & 13.62$\pm$0.02 & 12.92$\pm$0.02 & 12.51$\pm$0.01 & 12.13$\pm$0.04 & 10.77$\pm$0.02 &  3.00$\pm$0.04 &  5.82$\pm$0.04 &   6.80$\pm$0.08 &  0.74 \\
  15 &  4 & 15.56$\pm$0.05 & 14.89$\pm$0.04 & 14.51$\pm$0.03 & - & 10.56$\pm$0.03 &  3.96$\pm$0.16 &  2.38$\pm$0.02 &   4.16$\pm$0.04 &  0.47 \\
.. & .. & .. & .. & .. & .. & .. & .. & .. & .. & .. \\
\hline
\end{tabular}
\begin{list}{}{}
 \item[$^{\rm 1}$] The full table is available in electronic form at http://amiga.iaa.es. The columns correspond to (1): galaxy identification according to CIG catalog; (2): morphological type; (3): Galactic and K--corrected magnitude in the g--band; (4): Galactic and K--corrected magnitude in the r--band; (5): Galactic and K--corrected magnitude in the i--band; (6): Galactic and K--corrected magnitude in the K$_S$--band; (7): stellar mass; (8): half--light radius derived by {\tt SExtractor}; (9): S\'ersic index fitted by {\tt GALFIT}; (10): effective radius fitted by {\tt GALFIT} (11): semiaxes ratio fitted by {\tt GALFIT}. Values of radii and stellar masses are given in CDS and AMIGA VO interface for both the cosmology adopted here ($\rm H_0=70 \rm \ km \rm \ s^{-1} \rm \ Mpc^{-1}$) as for that used in the previous AMIGA studies ($\rm H_0=75 \rm \ km \rm \ s^{-1} \rm \ Mpc^{-1}$).
\end{list}
\end{center}
\end{table*}

\section{Data and sample selection}

This work is part of the AMIGA project \citep{2005A&A...436..443V}. Since its beginning, this project has made a refinement and multiwavelength characterization of the CIG. The data are being released and periodically updated at http://amiga.iaa.es, where a Virtual Observatory compliant web interface with different query modes has been implemented. We applied the same sample selection as in \citet{2012A&A...540A..47F}, considering both isolation and completeness criteria. The isolation criteria were defined in \citet{2007A&A...472..121V}, which reject galaxies with isolation parameters Q$>$-2 and $\eta_k$$>$2.4 \citep[the tidal strength created by all neighbors, $Q$, is more than 1$\%$ of the internal binding forces,][for the local number density, $\eta_k$, this translates into a value of 2.4]{atan1984} and with recession velocities V$_r$$<$1500 km/s (for lower values, the area searched for neighbors spreads too much on the sky). These conditions imply that the evolution of all galaxies in our selected 
sample is dominated by their intrinsic properties. There are 657 objects in the complete AMIGA sample that fulfill the above isolation criteria.

For the following analysis we downloaded the images of our galaxies from the SDSS--III \citep[Data Release 8, DR8][]{2011ApJS..193...29A} in g, r and i bands. A new approach for background subtraction was applied in DR8 that first models the brightest galaxies in each field so that the estimated sky background remains unaffected \citep{2011AJ....142...31B}. From the 657 AMIGA galaxies, we found that 497 were observed in the SDSS--DR8. In a few cases more than one frame was needed in order to fully reconstruct the image of a galaxy. Frames were combined using the iraf task {\tt imcombine}. Five galaxies were rejected because a bad combination of the images caused by a bad astrometry in some of the frames or because there is not adjacent image for combining. Through the direct analysis of the images, we found and removed 21 galaxies strongly affected by saturated stars. We also removed 16 galaxies with unknown redshifts because the redshift will be needed for the following analysis.

We masked the stars and derived the optical parameters of these galaxies using {\tt SExtractor} \citep{1996A&AS..117..393B} in the g, r and i bands. We chose the MAG$\_$AUTO in the catalogs, which provides a good approximation to the total magnitude of the objects. These magnitudes were corrected for Galactic dust extinction by applying the reddening corrections computed by SDSS following \citet{1998ApJ...500..525S}. We calculated the rest--frame magnitudes and the stellar masses by fitting the spectral energy distribution (SED) using the routine {\tt kcorrect} \citep{2007AJ....133..734B}, which assumes an initial mass function (IMF) of Chabrier. We used also Ks--band photometry from 2MASS \citep{2006AJ....131.1163S} for the SED fitting when available. In this step we rejected 3 galaxies because of an unrealistic stellar mass probably caused by an erroneous Ks--band magnitude. The final sample consists of 452 isolated galaxies. The radial velocities of these galaxies are mainly between 1500 and 15000 $\rm km 
\ s^{-1}$, with 10 galaxies having velocities up to 24000 $\rm km \ s^{-1}$. Galactic extinction and K--corrected magnitudes in each band, as well as stellar masses are presented in Table~\ref{dat}. Since {\tt kcorrect} assumes a cosmology with $\rm H_0=100 \rm \ km \rm \ s^{-1} \rm \ Mpc^{-1}$, the stellar masses were transformed to our cosmology as log (M$_{\ast} h^{-2}$), where h=H$_0$/100=0.7. Note that the magnitudes presented in Table~\ref{dat} are consistent with those presented in Table 2 of \citet{2012A&A...540A..47F} (1$\sigma$$\sim$0.1 mag), since there model magnitudes taken from SDSS catalog were used. 

The structural modeling of the galaxies was made by fitting a S\'ersic profile to the i--band SDSS images, using the {\tt GALFIT} package of \citet{2010AJ....139.2097P}. The model was convolved with a point--spread function (PSF) generated from the SDSS psField in each band, and we used the parameters derived by {\tt SExtractor} as inputs in the {\tt GALFIT} code. {\tt GALFIT} was designed to work in counts and it underestimates the galaxy size if the image has very low intensity values. To solve this problem, DR8--images pixel values were transformed from nanomaggies to counts, dividing by the NMYU (calibration value translating counts to nanomaggies) parameter in the header. The {\tt GALFIT} code provides the effective semimajor axis (a$_{\rm e}$) rather than the circular effective radius (R$_{\rm e}$). Since local works use structural parameters defined in a circular annuli, we used $ \rm R_e=a_e \ x \ \sqrt{b/a}$, where $b/a$ is the axis ratio given by {\tt GALFIT}, for a proper comparison with other 
samples. Physical sizes were calculated using the updated distances of the AMIGA galaxies presented in \citet{2012A&A...540A..47F}. Since these distances were calculated using $\rm H_0=75 \rm \ km \rm \ s^{-1} \rm \ Mpc^{-1}$, the physical sizes were changed to the cosmology adopted here as $\rm R_e \ x \ (0.75/0.7)$. Structural parameters derived using both methods are presented in Table~\ref{dat} (the complete table is available online).

To check the fits we compared the S\'ersic index ($n$) of each galaxy with its visual morphological classification \citep[][revised in Fern\'andez Lorenzo et al. 2012]{2006A&A...449..937S}, and the effective radius (R$_{\rm e}$) with the half--light radius obtained with {\tt SExtractor} (R$_{50}$). Our sample is composed by 385 late--type galaxies and 67 early--types. While 77\% of late--type galaxies have 0.5$<$n$<$2.5, only 22\% of early--types were fitted with 2.5$<$n$<$4.5. We found good agreement between R$_{\rm e}$ and R$_{50}$ for objects into these S\'ersic index ranges but large discrepancies for the rest of the sample, which indicates a bad fit for these galaxies.

\section{Stellar mass--size relation}

\begin{figure}
\centering
      \includegraphics[angle=0,width=8.2cm]{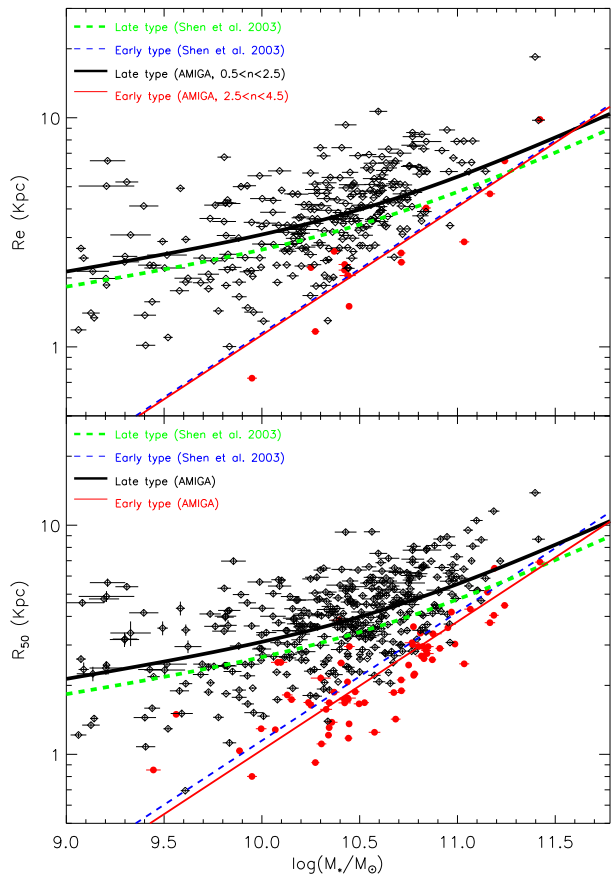}
      \caption{Stellar mass size relation for the AMIGA galaxies using the S\'ersic effective radius (top) and the half--light radius (bottom). Open black diamonds represent the AMIGA late--types (1$<$T$<$10) and solid red points represent the AMIGA early--type (-5$<$T$<$0) galaxies. In the top pannel, we consider only late--type objects with S\'ersic index 0.5$<$n$<$2.5 and early--types in the range 2.5$<$n$<$4.5. The blue and green dashed lines represent the relations of \citet{2003MNRAS.343..978S} for early and late--types, respectively. The solid lines represent the fits to AMIGA early-- (red) and late-- (black) type galaxies using the same functions as \citet{2003MNRAS.343..978S} (Eq. 1 and 2), but allowing a change only in the zero--point (equivalent to a change in the galaxy size).}
\label{fig1}
   \end{figure}

In Fig.~\ref{fig1}, we plot the size versus stellar mass values for our sample of galaxies. Our sample was separated into early (-5$<$T$<$0) and late--type (1$<$T$<$10) galaxies. The relation between stellar mass and galaxy size was established in the local universe by \citet{2003MNRAS.343..978S}, based on the Sloan Digital Sky Survey (SDSS). For comparison with the stellar mass--size relation obtained here for isolated galaxies, we overplotted the fits given by \citet{2003MNRAS.343..978S} for the SDSS galaxies (without any selection of environment), that provides the distribution of the S\'ersic half--light radius as a function of the stellar mass. For early--types, this function has the form:

\begin{equation}
\log {\rm R_{\rm e}(kpc)}=\log \rm{b} + a \log \left(\frac{M_{\ast}}{M_{\odot}}\right) ,
\end{equation}
where a=0.56 and b=2.88x10$^{-6}$, according to the relation fitted by \citet{2003MNRAS.343..978S}. In the case of late--type galaxies, the function used in the fit is:

\begin{equation}
\log {\rm R_{\rm e}(kpc)}=\log \rm \gamma + \alpha \log \left(\frac{M_{\ast}}{M_{\odot}}\right)+(\beta-\alpha)\log\left(1+\frac{M_{\ast}}{M_0}\right) ,
\end{equation}
where $\gamma$=0.1, $\alpha$=0.14, $\beta$=0.39, and M$_0$=3.98x10$^{10}$M$_{\odot}$ are the values fitted by \citet{2003MNRAS.343..978S}.

We have represented these relations using both effective ({\tt GALFIT}) and half--light ({\tt SExtractor}) radius as size estimators, because not all of our galaxies could be fitted by an accurate S\'ersic profile, as we have shown in the previous section. To check if there is some difference between AMIGA sample and \citet{2003MNRAS.343..978S} sample, we have fitted to AMIGA galaxies the same functions as \citet{2003MNRAS.343..978S} (Eq. 1 and 2), allowing a change only in the zero--point \citep[in the case of][zp=$\log$ b=-5.54 and zp=$\log\gamma$=-1 for the early and late--type galaxies respectively]{2003MNRAS.343..978S}, which can be interpreted as a change in the galaxy size. We found no difference in the stellar mass--size relation for early--type galaxies with respect to the \citet{2003MNRAS.343..978S} one when using the S\'ersic effective radius as size estimator, but a slight difference when using the half--light radius from SExtractor 
($\Delta$zp=$\log $R$_{50,AMIGA}-\log$ R$_{e,Shen}=-$0.04$\pm$0.02). However, the same difference in the zeropoint of $\Delta$zp=$\log $R$_{50,AMIGA}-\log$ R$_{e,Shen}=$0.07$\pm$0.01 is found for late--type galaxies when using both size estimators, which means that the late--type isolated galaxies would be $\sim$1.2 times larger or would have $\sim$0.56 times less stellar mass than similar less isolated objects. There are 12 galaxies in the AMIGA sample with stellar masses $\log$(M$_{\ast}$)$<$9 that were excluded from the fit since they can be considered as dwarf galaxies \citep{2012ApJ...757...85G}. In the later analysis we used the half--light radius obtained with {\tt SExtractor} as size estimator for all the AMIGA galaxies because it is independent of the fit given by {\tt GALFIT}, but consistent with the S\'ersic effective radius.

The result obtained in this work is in contrast with that of \citet{2006A&A...456...91G}, who concluded that isolated galaxies are smaller than objects in interaction. Luminosities and dynamical masses of both their samples of isolated and interacting galaxies were compared in \citet{2004A&A...420..873V}. They found no isolated galaxies with high mass, whereas no perturbed galaxies with low mass. Then, the discrepancy of \citet{2006A&A...456...91G} with our work could be caused by a different mass distribution between the sample of isolated and interacting galaxies they compared. However, this conclusion is based on dynamical masses and the comparison between stellar masses of their samples could be different. In \citet{2004A&A...420..873V}, they also compared the luminosity and size of isolated and interacting galaxies. They found that both of their samples satisfy the same luminosity--size relation. Since no stellar mass--size relation is given in this paper, it is difficult to know whether this represents 
a real difference with our result or it is a consequence of a different luminosity--stellar mass relation for isolated and interacting galaxies.

\subsection{The stellar mass--size relation in other environments}

The comparison between samples analyzed in a different way should be treated with special care as it can lead to wrong results. In the case of \citet{2003MNRAS.343..978S}, the sizes were calculated by fitting a S\'ersic profile to the galaxy in the z--band, and the criteria for morphological separation is based on S\'ersic index, color, and concentration. Several works have investigated the accuracy of these relations since they are widely used as comparison in high redshift studies. \citet{2010ApJ...712..226V} found systematically lower radii ($\sim$0.1 dex) in their cluster early-- and late--type galaxies with respect to the  \citet{2003MNRAS.343..978S} relations. However, this difference could be explained with the systematic offset in mass they find with respect to SDSS masses. Contrary to \citet{2010ApJ...712..226V}, \citet{2009MNRAS.398.1129G} claimed that the S\'ersic index, magnitude and effective radius derived by \citet{2005AJ....129.2562B} (the NYU--VAGC catalog) and used by 
\citet{2003MNRAS.343..978S} are underestimated. 

 \begin{figure}
\centering
      \includegraphics[angle=0,width=8.2cm]{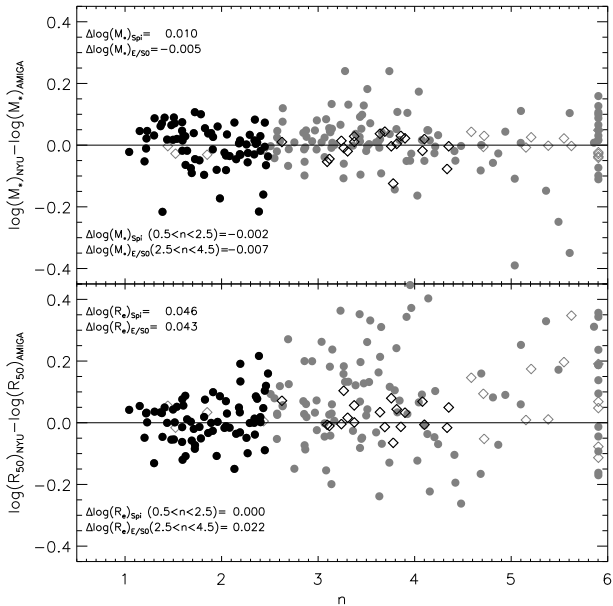}
      \caption{Comparison between AMIGA and NYU--VAGC stellar mass (top) and half--light radius (bottom) as function of the NYU--VAGC S\'ersic index for visually classified late--type galaxies (solid points) and early--types (open diamonds). Late--types with 0.5$<$n$<$2.5 and early--types with 2.5$<$n$<$4.5 are represented in black, while the rest of objects are represented in grey.}
\label{fig2}
   \end{figure}

To check the reliability of our results, we used the data of the NYU--VAGC catalog. They provide S\'ersic fits to the radial profile of each galaxy in the i--band, and the NYU--VAGC stellar masses were computed using {\tt kcorrect} as our stellar masses. In Fig.~\ref{fig2}, we compare our stellar masses and half--light radius with the NYU--VAGC data for the AMIGA galaxies in common. Our stellar masses are in good agreement with the NYU--VAGC ones. In the cases of effective radius, we found better agreement when considering only late--type galaxies with a S\'ersic index 0.5$<$n$<$2.5 and early--types in the range 2.5$<$n$<$4.5, as we found by fitting our galaxies with {\tt GALFIT} S\'ersic profiles. However, a large number of late--types (60\%) are fitted with a S\'ersic index higher than 2.5, which means that without a morphological visual classification, these objects would be considered as E/S0. We found a very similar result by fitting the stellar mass--size relation of this subsample of spiral 
($\Delta$zp=0.018$\pm$0.018) and early--type ($\Delta$zp=-0.006$\pm$0.021) galaxies using our data and the NYU--VAGC one. However, the zeropoint obtained for this subsample of late--type galaxies is closer to the \citet{2003MNRAS.343..978S} relation than the one found for our whole sample. We investigated the reason of this discrepancy and found that the spirals fitted with S\'ersic index greater than 2.5 are the most massive and earliest. This means that the environmental dependence found previously in this work is most important for the most massive galaxies. However, the AMIGA sample has the characteristic of being composed mainly of late--type galaxies so we need to investigate the stellar mass--size relation for each morphological type. 

\subsection{The stellar mass--size relation as function of the morphological type.}

Since, as explained above, we need a sample with visual determination of the morphologies, we used the sample of \citet{2010ApJS..186..427N}, which includes detailed visual morphological classifications for 14,034 galaxies in the SDSS DR4. We selected only objects with available morphological classification and redshift 0.01$<$z$<$0.05 (8976) to better match our AMIGA sample  (98$\%$ of our galaxies are in this redshift range). We also imposed a cut in magnitude in the r--band of mag$_r<$14.5, which roughly correspond to our completeness limit in the B--band (magB$<$15.3). For the stellar masses and structural parameters we used the data of the NYU--VAGC catalog. We imposed a S\'ersic index cut of 0.5$<$n$<$2.5 for late--type galaxies and 2.5$<$n$<$4.5 for early--types. The final sample of comparison is composed by 353 late-- and 824 early--type galaxies. We fitted the same function as \citet{2003MNRAS.343..978S}, allowing a change only in the zero--point (which can be interpreted as a change in galaxy size),
 as we did in Sect. 3 for the AMIGA galaxies. We found good agreement between the \citet{2010ApJS..186..427N} sample and the local relations of \citet{2003MNRAS.343..978S} for both early-- and late--type galaxies. 

 \begin{figure*}
\centering
      \includegraphics[angle=0,width=17.5cm]{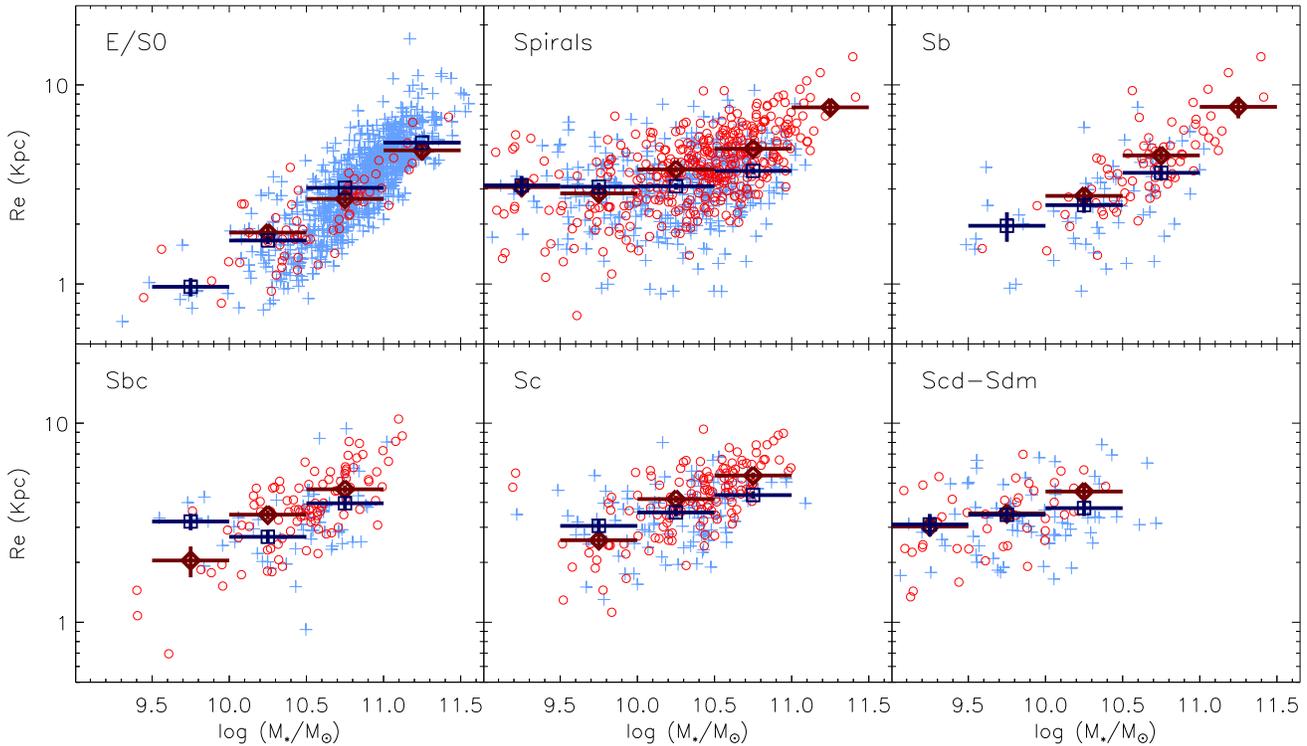}
      \caption{Stellar--mass size relation for galaxies in the AMIGA (red points) and the \citet{2010ApJS..186..427N} (blue crosses) samples. The mean half--light radius in each mass bin (represented by the error bars) for the AMIGA (red diamonds) and the \citet{2010ApJS..186..427N} samples (blue squares), are shown for each morphological type. The $\overline{R}e$ error bars represent the standard error in the mean in each case. The mean half--light radius was computed only for bins with more than 5 galaxies.}
\label{fig3}
   \end{figure*}

\begin{table*}
\caption{Mean size values as function of the stellar mass for AMIGA and the \citet{2010ApJS..186..427N} sample (see Sect.3.2).}
\label{col}
\begin{center}
\begin{tabular}{rccccccccc}
\hline
\tiny
\\
\normalsize
{\bf $\log$(M$_{\ast}$)} & {\bf $\overline{R}e$(AMIGA)} & {\bf $\overline{R}e$(Nair)} & p(K--S) & {\bf $\overline{R}e$(AMIGA)} & {\bf $\overline{R}e$(Nair)} & p(K--S) & {\bf $\overline{R}e$(AMIGA)} & {\bf $\overline{R}e$(Nair)} & p(K--S) \\
\hline
{} & \multicolumn{3}{c}{\bf E/S0} & \multicolumn{3}{c}{\bf Spirals} & \multicolumn{3}{c}{\bf Sb}\\
  $[$9-9.5$]$ &  - &  - &  - &  3.07$\pm$0.32 &  3.13$\pm$0.35 &  0.89 &  - &  - & - \\
 $[$9.5-10$]$ &  - &  0.97$\pm$0.10 &  - &  2.85$\pm$0.18 &  3.08$\pm$0.15 &  0.12 &  - &  1.96$\pm$0.33 &  - \\
$[$10-10.5$]$ &  1.81$\pm$0.12 &  1.65$\pm$0.07 &  0.31 &  3.76$\pm$0.11 &  3.10$\pm$0.11 &  0.00 &  2.77$\pm$0.20 &  2.49$\pm$0.21 &  0.05 \\
$[$10.5-11$]$ &  2.68$\pm$0.14 &  3.04$\pm$0.05 &  0.05 &  4.78$\pm$0.12 &  3.70$\pm$0.13 &  0.00 &  4.42$\pm$0.24 &  3.62$\pm$0.32 &  0.02 \\
$[$11-11.5$]$ &  4.69$\pm$0.51 &  5.13$\pm$0.11 &  0.02 &  7.71$\pm$0.65 &  - &  - &  7.76$\pm$0.97 &  - &  - \\
{} & \multicolumn{3}{c}{\bf Sbc} & \multicolumn{3}{c}{\bf Sc} & \multicolumn{3}{c}{\bf Scd-Sdm}\\
  $[$9-9.5$]$ &  - &  - & - &  - &  - &  - &  3.03$\pm$0.31 &  3.10$\pm$0.40 &  0.92 \\
 $[$9.5-10$]$ &  2.04$\pm$0.36 &  3.21$\pm$0.31 &  0.03 &  2.58$\pm$0.25 &  3.05$\pm$0.21 &  0.01 &  3.52$\pm$0.30 &  3.47$\pm$0.25 &  0.97 \\
$[$10-10.5$]$ &  3.47$\pm$0.18 &  2.69$\pm$0.19 &  0.10 &  4.16$\pm$0.16 &  3.56$\pm$0.16 &  0.02 &  4.54$\pm$0.30 &  3.74$\pm$0.33 &  0.00 \\
$[$10.5-11$]$ &  4.65$\pm$0.21 &  3.96$\pm$0.29 &  0.02 &  5.46$\pm$0.20 &  4.35$\pm$0.19 &  0.00 &  - &  - &  - \\
$[$11-11.5$]$ &  - &  - &  - &  - &  - &  - &  - &  - &  - \\
\hline
\end{tabular}
\end{center}
\end{table*}

To check the environmental dependence of the stellar mass--size relation for each Hubble type, the AMIGA and \citet{2010ApJS..186..427N} samples were divided into 6 morphological ranges:  the two main morphological groups: E/S0 (-5$<$T$<$0), and Spirals (1$<$T$<$8); and the four subgroups of spirals: Sb (T=3), Sbc (T=4), Sc (T=5), and Scd--Sdm (6$<$T$<$8). An independent subgroup of Sa--Sab galaxies was not considered because there are only 19 objects in the AMIGA sample with this morphological classification. The results are presented in Fig.~\ref{fig3}. To investigate a different environmental effect for high and low mass galaxies, we divided the stellar mass range into 5 bins. The mean half--light radius for each morphological type and stellar mass range was calculated when the bin was composed by more than 5 galaxies. These mean values and their mean standard errors (1$\sigma$) are also drawn in Fig.~\ref{fig3} and presented in Table.~\ref{col}. We find no significant difference at 3$\sigma$ level in the 
case of early--type galaxies, with two of three stellar mass ranges having differences $\leq$1$\sigma$. We also find no significant difference at 1$\sigma$ level for less massive spirals ($\log$(M$_{\ast}$)$<$10), while the most massive bins (10 $<\log$(M$_{\ast}$) $<$11) present a clear difference $>$3$\sigma$. Looking at different spiral types, we observed that the later is the morphological type, the larger is the galaxy size for a fixed stellar mass range. This means that a comparison between two samples with different spiral populations may lead to a wrong result. Then, a simple segregation between early--type and spiral galaxies would not be enough when comparing samples in different environments because their spiral population are probably different \citep[more late--types in low density environments, e.g.][]{2006A&A...449..937S}. We also observed that the difference in size found for the high mass bins is significant for almost all spiral morphological types. For each stellar mass bin and 
morphological type, we have also calculated the probability, given by the Kolmogorov--Smirnov two--sample test, that the galaxy size distribution of the \citet{2010ApJS..186..427N} sample is indistinguishable from the AMIGA one (p(K--S)$>$0.05). The values obtained from this statistical test, presented in Table~\ref{col}, are in agreement with and confirm the above mentioned differences in size. The same test was performed for the distribution of stellar masses inside each bin. The only bin that is not comparable in stellar mass (p$<$0.05) between Nair and AMIGA is that of E/S0 with 11$<\log$(M$_{\ast}$) $<$ 11.5, probably because the number of AMIGA E/S0 in this stellar mass range is small (8).

This lack of dependence on galaxy size with environment for spiral galaxies with stellar masses $\log$(M$_{\ast}$) $<$ 10 contrasts with the result of \citet{2010MNRAS.402..282M}, who found that cluster spirals with $\log$(M$_{\ast}$) $<$ 10 are smaller than similar field objects. In this work, \citet{2010MNRAS.402..282M} used the effective semimajor axis as size estimator (instead of the circularized radius), derived by fitting a S\'ersic model with {\tt GALFIT}. To compare with their results, we have calculated our mean size values using also the effective semimajor axis ($\overline{a}_e$) derived with {\tt GALFIT} for our galaxies. Since only 22\% of our early--types were fitted with 2.5$<$n$<$4.5, we only have enough objects to compare in their stellar mass bin of [10, 11.5]. Our result of $\overline{a}_e$=4.07$\pm$0.74 kpc agrees well with their value for elliptical galaxies (4.29$\pm$0.59 kpc), but we do not have enough objects to separate between elliptical and lenticular galaxies. In the case of 
spirals, we found a mean size of 4.24$\pm$0.46 kpc in the mass bin of [9,9.5], which is larger than their value of 3.14$\pm$0.16 kpc, presumably because we have mostly very late--type galaxies in this bin (Scd--Sdm). However, in the mass bin of [9.5,10] our result of 3.97$\pm$0.26 kpc is very similar to their mean value (4.00$\pm$0.18 kpc), and significantly larger than their mean size for cluster (3.42$\pm$0.12 kpc) and core cluster (3.52$\pm$0.39 kpc) spirals. Finally, in the mass bin of [10,11], we found a larger mean size of 5.81$\pm$0.15 kpc compared with their field (4.85$\pm$0.21 kpc), cluster (5.10$\pm$0.21 kpc) and core cluster (5.61$\pm$0.46 kpc) spiral galaxies. 

 \begin{figure*}
%\centering
      \includegraphics[angle=0,width=17.cm]{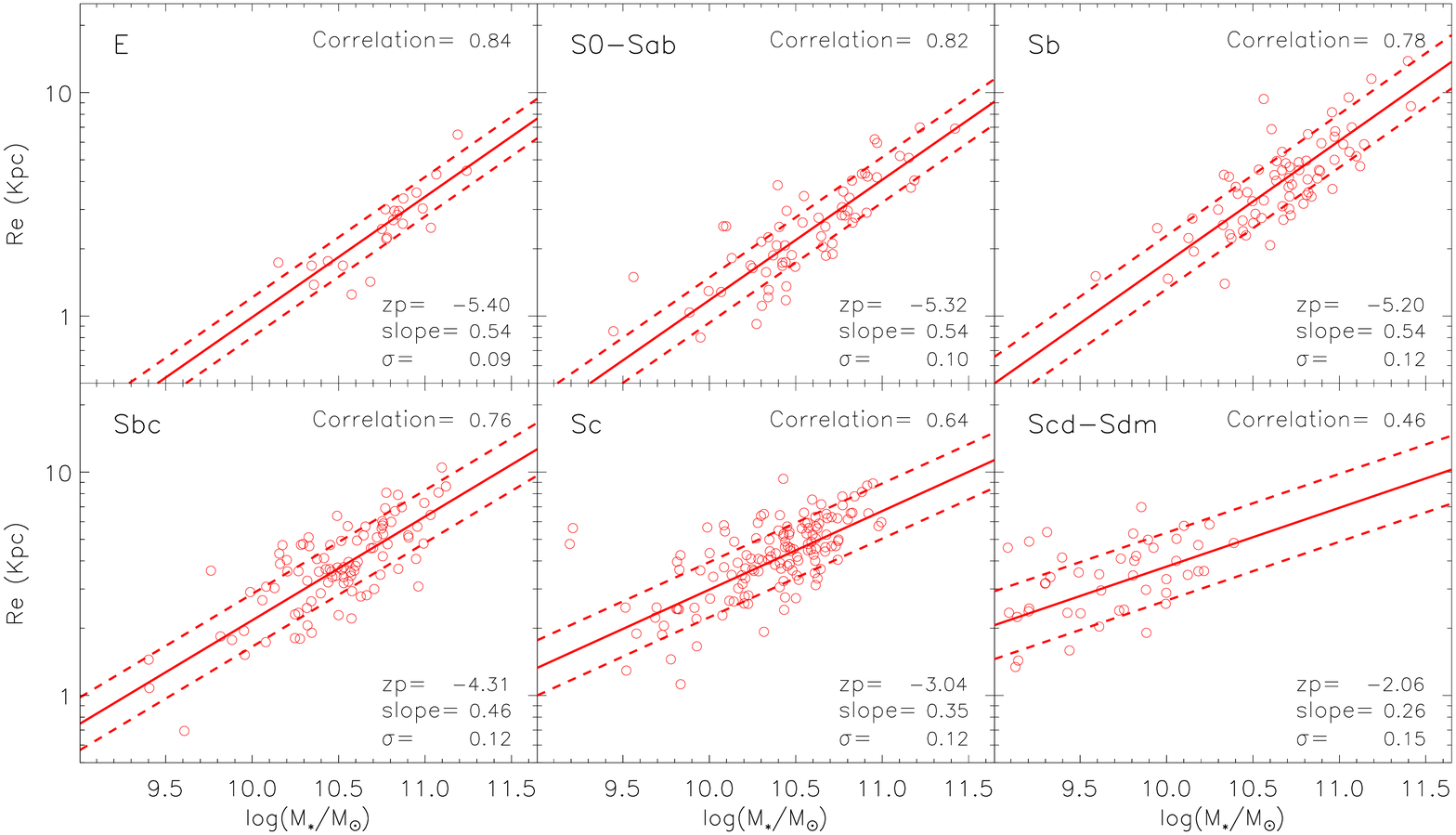}\\
      \includegraphics[angle=0,width=17.cm]{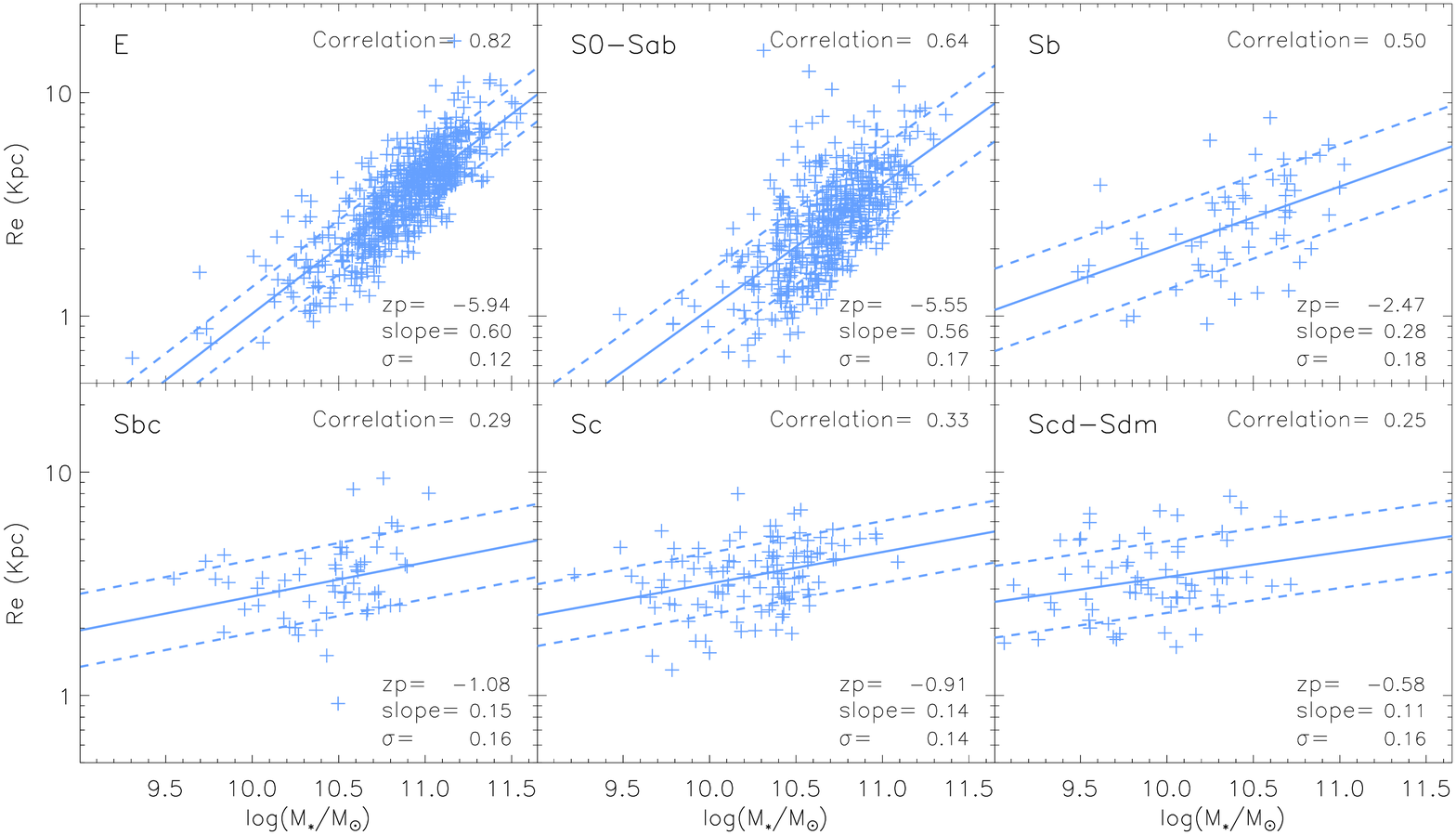}
      \caption{Stellar--mass size relation for galaxies in the AMIGA (red points) and the \citet{2010ApJS..186..427N} (blue crosses) samples. The solid and dashed lines represent the linear fit and its 1$\sigma$ confidence interval for galaxies in each panel. The zeropoint, slope, and sigma are given in each case, as well as the correlation coefficient.}
\label{fig4}
   \end{figure*}

Looking at spiral subtypes in Fig.~\ref{fig3}, we note a different trend in the stellar mass--size relation than that obtained when considering all spiral types together (Eq. 2). In Fig.~\ref{fig4} we fitted each morphological subtype with a linear function between log(R$_e$) and $\log$(M$_{\ast}$), the same used by \citet{2003MNRAS.343..978S} for early--type galaxies (Eq. 1). The same was done for the \citet{2010ApJS..186..427N} sample. In all cases, the relation is better defined and has less scatter for the isolated galaxies. Then, the AMIGA sample better clarifies the relations between fundamental parameters of galaxies because the blurring effects of environment are minimized. The slope obtained for elliptical galaxies is very similar to that of \citet{2003MNRAS.343..978S}, and remains constant up to Sb galaxies. Starting with Sb, the slope becomes less steep as we go to later types. This change in slope is probably caused by the changing bulge/disk ratio, and it is responsible for the function obtained 
when all spirals types are fitted together (Eq. 2). That function is strongly dependent on the percentage of each spiral type considered in the fit, and therefore, comparisons in different environments and redshifts should be done taking this in mind. 

\section{Discussion}

We find interesting differences between the stellar mass--size relation for very isolated galaxies and the Nair sample. 1) When we divide our sample into morphological subtypes we find less scatter and a better defined correlation between size and mass for late--type spirals (which represent 2/3 of our sample). Assuming that the two samples are different only in environmental density we suggest that the AMIGA sample provides a clearer view of the intrinsic physical relation between size and mass because the blurring effects of environment are minimized. This is likely true of all or most physical measures and correlations involving galaxies and has been seen since a larger scatter in colors was found for interacting galaxies \citep{1978ApJ...219...46L}. 2) As the morphological type becomes later the galaxy size for a fixed stellar mass range becomes larger. Also the slope of the stellar mass--size relation changes systematically across the spiral sequence becoming less steep for later types. The change in 
size and slope across the Sb--Sc isolated galaxy majority population (where mean galaxy luminosity remains constant) is probably caused by the changing bulge/disk ratio. 3) We find a difference between the stellar mass--size relations for AMIGAs and the Nair sample especially for high mass spirals 10$<$$\log$(M$_{\ast}$)$<$11 which are the dominant population in a sample of bright isolated galaxies. The difference can be interpreted in several ways: a) Isolated galaxies have systematically lower stellar masses. b) Isolated spiral galaxies could be larger in physical size than similar objects in denser environments. In the next subsections, we investigate possibilities a and b. c) Other explanations for the difference involve sample differences and differences in data processing. The former possibility includes systematic differences in galaxy classification. The latter has been checked and minimized through the comparison shown in Fig.~\ref{fig2}.

\subsection{Passive star formation history of isolated galaxies}

Are isolated galaxies less massive because they live in such low--density environments? One must consider their likely star formation history and accretion rate. In addition to internal processes, a galaxy can gain stellar mass by accretion of neighbors and by stimulation of star formation through interactions. Both are less likely to play an important role in very isolated galaxies where the cross section for accretion and interaction stimulation by neighbours are assume to be low. It is known that galaxy interactions and the presence of companions are associated with enhanced star formation \citep{1978ApJ...219...46L,2003MNRAS.346.1189L}. \citet{1978ApJ...219...46L} were first to demonstrate an increased dispersion in B--V colors, and an excess of bluer colors among interacting galaxies. This enhancement is stronger if the luminosities of companion and host are similar \citep{2006AJ....132..197W,2012MNRAS.426..549S}. AMIGA galaxies were selected with an isolation criterion that increases the probability 
that they have had few or no major interactions in at least the last 3 Gyr. Results obtained in previous AMIGA studies \citep[lower L$_{FIR}$, colder dust temperature, less molecular gas, redder colors, etc.][]{2007A&A...462..507L,2011A&A...534A.102L,2012A&A...540A..47F} point to a lower star formation rate compared to galaxies in denser environments. Assuming that the star formation history of these galaxies has been passive during most of their live leads to the expectation that they may have less stellar mass than galaxies in denser environments. This may be a good qualitative explanation, however we need to quantify whether the stellar mass increase is due to SF enhancement caused by interaction.

\citet{2011A&A...534A.102L} calculated an average star formation rate of 0.7·M$_{\ast}$ yr$^{-1}$ for the Sb--Sc isolated galaxies. On the other hand, \citet{2012MNRAS.426..549S} derived an average enhancement of 1.9 in the SFR of galaxies in close pairs. Considering this difference in the SFR during 3Gyr, we find that an isolated galaxy with $\log$(M$_{\ast}$)=10.5 (the average value for the Sb--Sc spirals) would only show a 5\% stellar mass deficit relative to a galaxy that suffered an interaction. In Section 3, we estimated that our isolated spirals show $\sim$44\% less stellar mass than galaxies in denser environments in order to explain the discrepancy with the relation of \citet{2003MNRAS.343..978S}. An enhancement in the SFR caused by interaction with close companions cannot account for a strong difference in stellar mass.

This is a simple test taking into account only one interaction and only in the last 3Gyr, while the life of a galaxy is much more complicated. Perhaps 10\% of the Nair sample might involve galaxies in pairs since \citet{1991ApJ...374..407X} found that 10\% of field galaxies are in pairs, while the number of pairs in AMIGAs is effectively zero. The main environmental difference between the two samples involves the local surface density of neighbors. Is the surface density around Nair galaxies high enough, and are interactions in loose groups frequent enough, to explain a mass difference due to interaction induced star formation? Nevertheless, from the information we have, we only can conclude that the difference in the stellar mass--size relation is not mainly caused by a different star formation history of objects in low and dense environments.

\subsection{Growth in size of isolated galaxies}
 
Evolution in the stellar mass--size relation is usually attributed to a growth in size of galaxies caused by minor mergers \citep{2005ApJ...625...23B,2005AJ....130.2647V}. Unlike the major merger rate which is higher in high density environments, the rate of minor mergers might depend on the initial local density. In the case that galaxies in low density environments had formed with a lower number of small companions, then these galaxies should have grown less than those in denser environments. On the other hand, if small companions are remnants of the galaxy formation process, an environmental dependence should not be expected for the growth in size of galaxies. We find in this study that isolated galaxies have grown at the same rate as galaxies in other environments, with massive spirals growing the most. Whatever the reason for the growth in size, it reflects that the same evolution has affected all galaxies. If we accept a role for minor mergers to explain the general growth of galaxies, then the small 
objects that have been accreted during the life of a galaxy should be remnants of the formation of that individual galaxy. In addition, the fact that isolated late--type galaxies in our sample are systematically larger than similar mass objects in other environments indicates that there is another environmental dependent process causing the larger sizes of isolated spirals. One possible explanation is that spiral galaxies in low density environments are the norm while such extended disks do not survive in higher density environments  \citep{2010MNRAS.402..282M}.

The criteria used by \citet{1973AISAO...8....3K} for selecting the sample have two effects in our galaxies. On one hand, they do not have major companions at large distances, minimizing effects of environment at large scale (tidal interactions, major mergers, etc). On the other hand, the number of small companions at close distances (satellites) until 4 magnitudes of difference with the CIG galaxy is also minimized, implying that our sample presents a deficit of local neighbors with respect to other samples. Then, another possibility for the larger sizes found in this work would be that AMIGA galaxies have accreted more satellite galaxies and as a consequence of this local accretion have created a local deficit of satellites.

In order to check this option we studied the influence of the local environment, calculating the number of satellites in a field of projected radius R=250 kpc \citep[using as criterion for defining satellite, the distance at which the 80\% of Milky Way satellites are found][]{2012MNRAS.427.1769F}, for the 207 AMIGA spiral galaxies with more than 80$\%$ spectroscopic completeness in the SDSS DR8. Since the SDSS spectroscopic database is complete until r$\sim$17.5 and our sample, until r$\sim$14.5, we consider only objects: 1) within 3 magnitude difference with respect to the central galaxy, and 2) with a difference in the recession velocity less than 1000 km/s. The same calculation was made for 336 spirals of \citet{2010ApJS..186..427N} that have more than 80$\%$ completeness in SDSS DR8. We find that 6\% of AMIGA spirals have one satellite (and one galaxy with two satellites). In contrast 43\% of the Nair sample show a satellite (19\% have two or more). We calculated again (see Table~\ref{col}) the mean size 
for Nair spirals in the stellar mass ranges [10,10.5] and [10.5,11] considering only galaxies without satellites, galaxies with no or one satellite and galaxies with two or more satellites. The mean size for spirals without satellites in the \citet{2010ApJS..186..427N} sample is very similar to the value obtained in Sect. 3.2: $\overline{R}e$[10--10.5]=3.18$\pm$0.16 kpc and $\overline{R}e$[10.5--11]=3.63$\pm$0.16 kpc. The mean size considering also galaxies with one satellite, is also consistent with the previous value. However, the mean size for spirals with two or more satellites is lower than the value obtained for the \citet{2010ApJS..186..427N} sample in Sect. 3.2: $\overline{R}e$[10--10.5]=2.87$\pm$0.22 kpc and $\overline{R}e$[10.5--11]=3.35$\pm$0.33 kpc.

 \begin{figure}
\centering
      \includegraphics[angle=0,width=8.2cm]{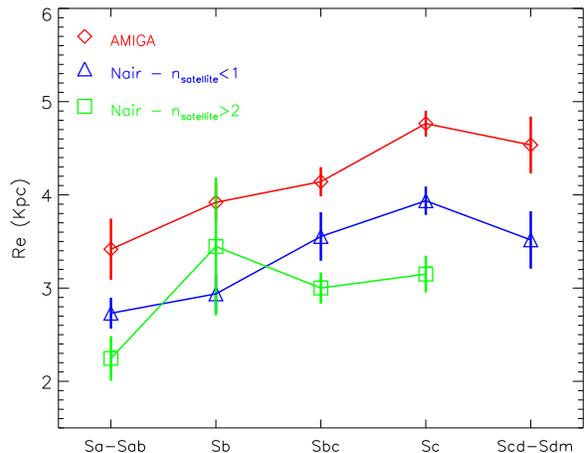}
      \caption{Mean size for galaxies in the stellar mass range [10--11] as function of each morphological spiral type, for the AMIGA sample (red diamonds), the galaxies in the \citet{2010ApJS..186..427N} sample having zero or one satellite (blue triangles), and galaxies in the \citet{2010ApJS..186..427N} sample with two or more satellites (green squares).}
\label{fig5}
   \end{figure}

In the case of individual spiral subtypes, we have calculated the mean size value for galaxies in the stellar mass range [10--11] for increasing the statistics. The same calculation was performed for the AMIGA sample, the \citet{2010ApJS..186..427N} sample of galaxies having zero or one satellite, and the \citet{2010ApJS..186..427N} galaxies with 2 or more satellites. The results are presented in Fig.~\ref{fig5}. In the three cases we find that spiral galaxies become larger as they become later types. Galaxies in the \citet{2010ApJS..186..427N} sample with no or one satellite are larger than those with two or more satellites for almost all morphological types. In all cases, the mean size of our isolated galaxies remains even larger than objects without satellites in the \citet{2010ApJS..186..427N} sample. 

These results confirm that the local environment is affecting the growth in size of galaxies. However, the objects without satellites in the \citet{2010ApJS..186..427N} sample are still smaller than our isolated galaxies. Then, another effect of the environment is expected to be the cause of the difference in size, and a truncation of the extended disks caused by effects of the large scale environment (group, cluster, etc) could be the reason \citep{2010MNRAS.402..282M}. In this sense, a study of the outer profiles of disks in different environments would be essential for disentangling the mechanisms involved in the growth in size of galaxies.

\section{Summary and conclusions}

We have investigated the stellar mass--size relation for a sample of isolated galaxies. This sample was selected from the AMIGA sample according to its completeness and the isolation criteria defined in \citet{2007A&A...472..121V}, which ensure that the tidal strength created by all neighbors is less than 1$\%$ of the internal binding forces. The stellar masses were derived by fitting the SED to the g, r, i and Ks--band photometry with {\tt kcorrect}. We used two different size estimations, the half--light radius obtained with {\tt SExtractor} and the effective radius calculated by fitting a S\'ersic profile to the i--band image of each galaxy with {\tt GALFIT}. We found a good agreement between both size estimations when the S\'ersic index given by {\tt GALFIT} was between 2.5$<$n$<$4.5 for early--types and between 0.5$<$n$<$2.5 for late--type galaxies.

The sample was divided in early (-5$<$T$<$0) and late--type galaxies (1$<$T$<$10) and both stellar mass--size relations were fitted using the same functions as in \citet{2003MNRAS.343..978S}, allowing a change only in the zero--point. We found no difference in the stellar mass--size relation for early--type galaxies with respect to the \citet{2003MNRAS.343..978S} one when using the S\'ersic effective radius as size estimator, but a slight difference when using the half--light radius from SExtractor ($\Delta$zp = $\log $R$_{50,AMIGA}-\log$ R$_{e,Shen}=-$0.04$\pm$0.02). For late--type galaxies, we found a difference in the zeropoint of $\Delta$zp = $\log $R$_{50,AMIGA}-\log$ R$_{e,Shen}=$ 0.07$\pm$0.01 independently of the size estimator used. This difference in the zero--point implies that the late--type isolated galaxies would be, on average, $\sim$1.2 times larger or would have $\sim$44\% less stellar mass than spiral galaxies in denser environments. 

To check the environmental dependence of the stellar mass--size relation for each Hubble spiral subtype, we compared our data with the sample of \citet{2010ApJS..186..427N}, which has no selection of environment and visually classified morphologies (necessary due to the differences between visual and S\'ersic classifications). The stellar mass and effective radius of the \citet{2010ApJS..186..427N} galaxies were taken from the NYU--VAGC catalog, after verifying that the NYU--VAGC data were consistent with those calculated by us for the AMIGA galaxies in common. Both samples were divided into six morphological ranges: E/S0 (-5$<$T$<$0), Spirals (1$<$T$<$8), Sb (T=3), Sbc (T=4), Sc (T=5), and Scd--Sdm (6$<$T$<$8). To investigate a different environmental effect for galaxies of high and low mass, we also divided the stellar mass range into five stellar mass bins between $\log$(M$_{\ast}$)=[9,11.5]. The main results of this comparison includes:

\begin{itemize}

\item There is no significant difference at 3$\sigma$ level in the case of early--type galaxies, with two stellar mass ranges having differences $\leq$1$\sigma$.

\item In the case of spiral types, the later is the morphological type, the larger is the galaxy size for a fixed stellar mass range. Therefore, if segregation of morphological spiral types is not done, the comparison of the stellar mass--size relation between samples of spiral galaxies in different environments may lead to a wrong result, since their spiral populations are probably different.

\item In the case of the less massive spirals ($\log$(M$_{\ast}$)$<$10), no significant difference is found at 1$\sigma$ level for AMIGA galaxies compared with similar objects in denser environments. This is in contrast with the result found by \citet{2010MNRAS.402..282M} of larger sizes for spiral galaxies in the field than in the cluster environment, but they did not perform a segregation by spiral subtype.

\item The high--mass AMIGA spirals (10 $<\log$(M$_{\ast}$) $<$ 11) present a clear difference $>$3$\sigma$ when comparing with the \citet{2010ApJS..186..427N} sample. This difference is found for all spiral types in the sense that they are larger than objects in denser environments. The difference in size is also significative when comparing with the cluster sample of \citet{2010MNRAS.402..282M}.

\item We find less scatter and a better defined correlation between size and mass for late--type spirals when we break the samples into morphological subtypes. Also the slope of the stellar mass--size relation changes systematically across the spiral sequence becoming less steep for later types.

\item The number of satellites around a galaxy affects its size. The galaxies in the \citet{2010ApJS..186..427N} sample with zero or one satellite have larger sizes than galaxies having 2 or more satellites. In all cases, the mean size of our isolated galaxies remains even larger than objects without satellites.

\end{itemize} 

The difference in the stellar mass--size relation for high mass spirals (10$<$$\log$(M$_{\ast}$)$<$11) found in this paper can be interpreted as a lower stellar mass or as a larger size for isolated galaxies comparing with similar objects in denser environments. We rejected the first explanation since the increase in the SFR caused by an interaction cannot explain the difference in stellar mass found here. Our results suggest that the environment plays a role in the growth in size of spiral galaxies, but not in the case of early--types. 

\section*{Acknowledgements}

This work has been supported by Grant AYA2011-30491-C02-01 co-financed by MICINN and FEDER funds, and the Junta de Andaluc\'ia (Spain) grants P08-FQM-4205 and TIC-114. We are grateful to the AMIGA team for their comments and suggestions. We thank Dr. Adriana Durbala for her help with the SDSS images.

Funding for SDSS-III has been provided by the Alfred P. Sloan Foundation, the Participating Institutions, the National Science Foundation, and the U.S. Department of Energy. The SDSS-III web site is http://www.sdss3.org/.

SDSS-III is managed by the Astrophysical Research Consortium for the Participating Institutions of the SDSS-III Collaboration including the University of Arizona, the Brazilian Participation Group, Brookhaven National Laboratory, University of Cambridge, University of Florida, the French Participation Group, the German Participation Group, the Instituto de Astrofisica de Canarias, the Michigan State/Notre Dame/JINA Participation Group, Johns Hopkins University, Lawrence Berkeley National Laboratory, Max Planck Institute for Astrophysics, New Mexico State University, New York University, Ohio State University, Pennsylvania State University, University of Portsmouth, Princeton University, the Spanish Participation Group, University of Tokyo, University of Utah, Vanderbilt University, University of Virginia, University of Washington, and Yale University.
 
We thank the SAO/NASA Astrophysics Data System (ADS) that is always so useful.

\end{document}